\documentclass{PoS}
\let\ifpdf\relax

\title{AGN Forecasts for the Cherenkov Telescope Array}

\ShortTitle{AGN forecasts for CTA}

\author{T. Hassan\\
        Universidad Complutense de Madrid, Spain\\
        E-mail: \email{thassan@gae.ucm.es}}

\author{N. Mirabal\\
        Universidad Complutense de Madrid, Spain\\
        E-mail: \email{mirabal@gae.ucm.es}}

\author{J. L. Contreras\\
        Universidad Complutense de Madrid, Spain\\
        E-mail: \email{contrera@gae.ucm.es}}

\author{for the CTA Consortium}

\abstract{ The First Fermi-LAT catalog (1FGL) represents the most
 complete list of sources in the GeV sky to date. We use the reported
1FGL spectral parameters to 
 extrapolate {\it Fermi} AGN spectra to the very-high
energy (VHE) range (15 GeV -- 300 TeV). 
The extrapolated VHE spectra are
then attenuated using  current estimations
 of the extragalactic background light (EBL) absorption as
a function of redshift. 
Using the expected effective areas and background rates of the
Cherenkov Telescope Array (CTA) from Monte Carlo simulations, 
we make a first order prediction of the AGN population
accessible to CTA in the VHE sky. We find that CTA should easily
triple the AGN detection rate of current ground-based Cherenkov 
telescopes. In addition, 
CTA will allow unprecedented access to high-redshift blazars out to
$z \approx 2$, and hence will start to reveal the EBL shape with
gamma-ray observations.}

\FullConference{AGN Physics in the CTA Era - AGN2011,\\
		May 16-17, 2011\\
		Toulouse, France}

\begin{document}

\section{Introduction}

The future Cherenkov Telescope Array (CTA) represents the next
generation of ground-based Cherenkov detectors \cite{design}. 
When completed, CTA 
is expected to improve
 the
sensitivity of present observatories such as H.E.S.S., MAGIC or VERITAS by an
order of magnitude. It will also expand the energy range coverage
from some tens of GeV to hundreds of TeV, opening a new window in
the Very High Energy (VHE) domain never reached with such exquisite 
detail. Furthermore, the 
synergy between the Large Area Telescope (LAT) on 
board of the {\it Fermi} Gamma-ray
Space Telescope and CTA  will allow nearly seamless 
coverage from MeV to TeV.

The CTA Observatory will consist of two arrays, one in each hemisphere.
The Southern hemisphere array is expected to be mainly 
dedicated to Galactic sources and bright active galactic nuclei (AGN),
whereas the Northern one will complement the Southern one, focusing
on northern extragalactic objects including AGN, galaxy clusters,  
gamma-ray bursts, and starburst galaxies. 
CTA is a complex project; as a result understanding its capabilities 
and limitations is not a simple
task. One possible way to evaluate its science impact is to simulated a 
population study using real data from
known gamma-ray sources expected to emit in the VHE range. 

Specifically, the {\it Fermi} Gamma-ray Space Telescope
provides an ideal set of candidates for this study 
through the First Fermi-LAT catalog (1FGL).
With 11 months of accumulated data, the 1FGL catalog contains  1451 sources 
characterized in the
100 MeV to 100 GeV energy range \cite{fermi}. Here, we exploit 
the overlap of the high energy
end of {\it Fermi} with the low energy range of CTA to attempt a
first order approximation of the extragalactic CTA sky. It is
 likely that many of the 1FGL catalog 
sources have no yet been discovered in VHE due to a lack of sensitivity of 
existing instruments,
but could be accessible to CTA. In fact, 39 out of the 45 VHE AGN
detected by ground-based Cherenkov observatories are found
in the 1FGL (B. Lott, 
priv. comm.). Therefore an extrapolation of  the 1FGL 
data to higher energies seems a sensible step to build a mock 
catalog of CTA sources. In this work we present this approximation
to forecast the AGN population for CTA.
 
Throughout this work, we rely on the CTA design concepts 
summarized in \cite{design}, where a number of
of possible array configurations including their 
effective areas and predicted
cosmic ray backgrounds are presented \cite{Konrad}. 
The proposed  configurations are composed 
mainly of 3 types of telescopes: \emph{large} (23 m diameter), 
\emph{medium} (around 12 m)
and \emph{small} (6-7m). Apart from number of telescopes, 
the individual configurations 
differ in other parameters such as the field of view or pixel
size \cite{design}. For simplicity, in this paper
we only consider candidate array E, that 
achieves a well balanced sensitivity over the full energy range
of CTA.

\section{Forecasting model}

The motivation for this work is to provide a method for 
estimating the number of AGN accessible  to CTA, 
based on the sources listed in the 1FGL catalog. The specific 
steps taken in determining 
the significance of each source are summarized 
in the following subsections.\\

\subsection{Selection criteria}

From the \emph{Fermi} AGN catalog \cite{Fermi_AGN_cat}, 
we first selected sources with counterparts in at least 
one of the commonly used 
AGN catalogs (CRATES, CGRaBS \cite{CGRaBS} or Roma-BZCAT \cite{Roma-BZCAT}).
Out of 671 sources we only select AGN with a measured 
redshift, a condition needed to apply the corresponding extragalactic background 
light (EBL) absorption, which is critical 
for AGN flux estimation in the VHE range. This results in 432
 sources. Finally, we discard sources whose spectra show a 
high curvature index C $>$ 11.34 \cite{fermi}, 
ending with a subset of 400 {\it Fermi} AGN including 247 flat-spectrum
radio quasars (FSRQs), 128 BL Lacs, and 25 of other/unknown type.

In order to extrapolate the AGN spectra to higher energies, 
we use the integral flux from 1 to 100 GeV in
ph cm$^{âˆ-2}$ s$^{âˆ-1}$ units (F1000)  
and spectral index $(\Gamma)$ furnished 
by the 1FGL catalog. For 
nearby hard sources $\Gamma<2$, a straight extrapolation 
could create runaway integrations, 
therefore we apply an artificial broken power law with 
a $\Gamma$ = 2.5 starting at 100 GeV to soften such spectra. The 
latter is in agreement 
with observed spectral properties \cite{zhang}.

\begin{figure*}[t]
  \centering
  \includegraphics[width=5in]{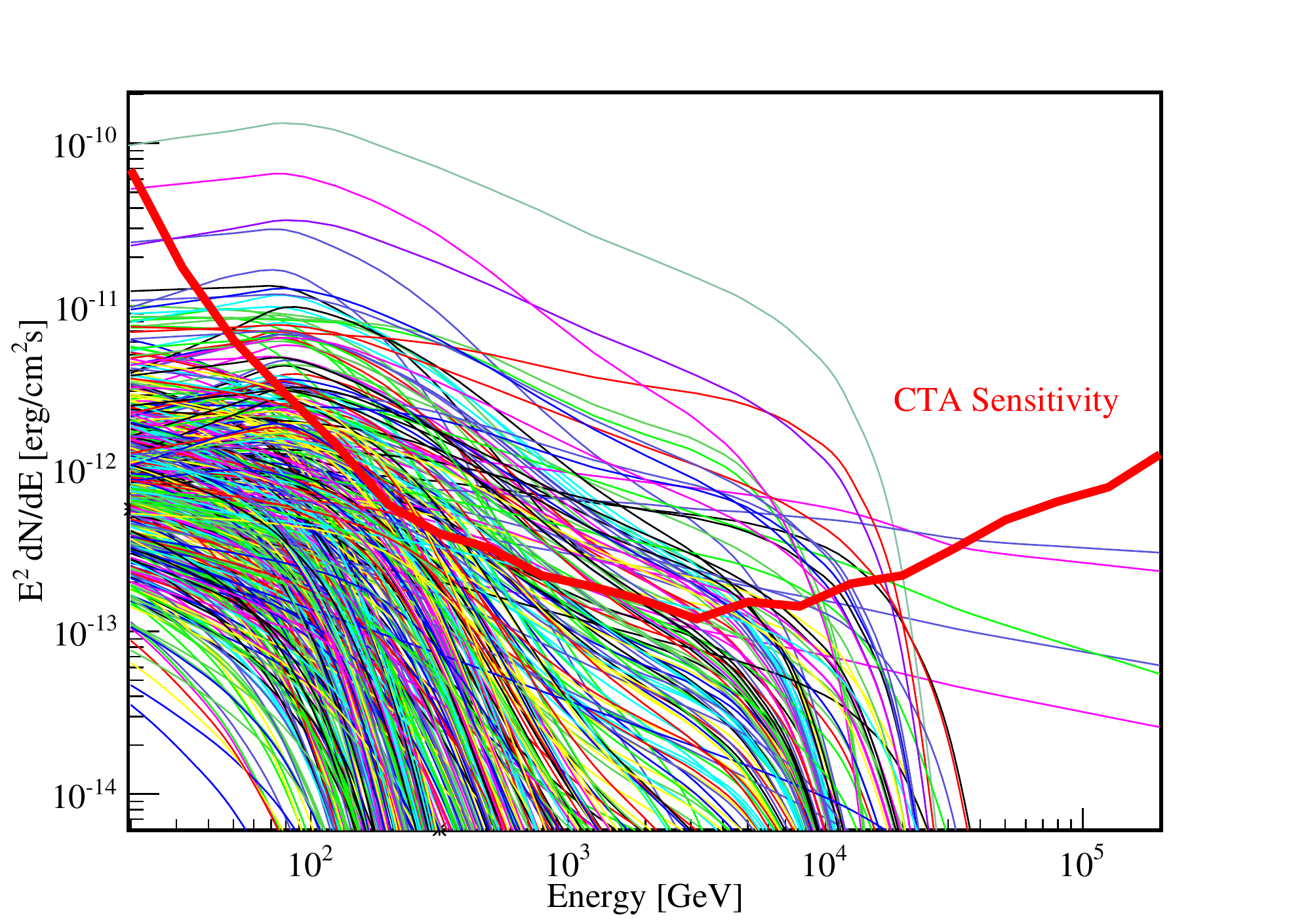}
  \caption{Extrapolated spectra from the 1FGL catalog attenuated with
the EBL model by \cite{EBL_Franceschini}. The thick red line marks the
integral sensitivity for a Crab-like spectrum 
expected for 50 hours at a zenith angle of 20$^{\circ}$ with
CTA candidate configuration E \cite{design}.}
  \label{spectra}
\end{figure*}

As commented above, the flux attenuation due to
infrared photons from the EBL is a critical factor that must be 
carefully taken into account for a precise and realistic 
flux calculation at energies above about 30 GeV. This effect produces
 a significant attenuation in photons with energies above 
tens of GeV through space \cite{Gould}. The observed 
{\it Fermi} spectra are thought to be free of EBL attenuation. 
However, for a proper extrapolation 
we applied the EBL model by \cite{EBL_Franceschini} through 
the whole range 
of CTA energies. The resulting
set of attenuated differential 
spectra can be seen on Figure \ref{spectra}.

\subsection{Significance Estimation}
Using the final extrapolated AGN spectra, we integrate the flux per energy bin 
weighted with the effective areas at a zenith angle of 20$^{\circ}$
for CTA candidate array E, 
obtained by the CTA Monte Carlo Work Package. The  
expression is subsequently
 multiplied by the
observation time (throughout this work we have assumed 50 hours) producing 
the total number of detected source photons. Total background rates 
for candidate array E are gathered from 
Monte Carlo simulations \cite{Konrad}.

For each source, the significance was calculated using Equation 
17 in \cite{Li&Ma} assuming $N_{on}$ (on region) to be the 
number of source photons plus the number of photons from the 
background (BG),  and $N_{off}$ fixed at the BG rate (off-region). The number 
$\alpha$ is given by the ratio of the sizes of the two regions, the 
ratio of the exposure times and the respective acceptances. For 
simplicity, an energy threshold of 20 GeV, 
5 off-regions for each on-region observations and a 5\% 
systematic error were considered in  this study \cite{design}.
A detection must exceed a significance above $5\sigma$ in 50 hours \emph{and} 
a signal over 5\% of the background.

\begin{figure*}[t]
  \centering
  \includegraphics[width=5in]{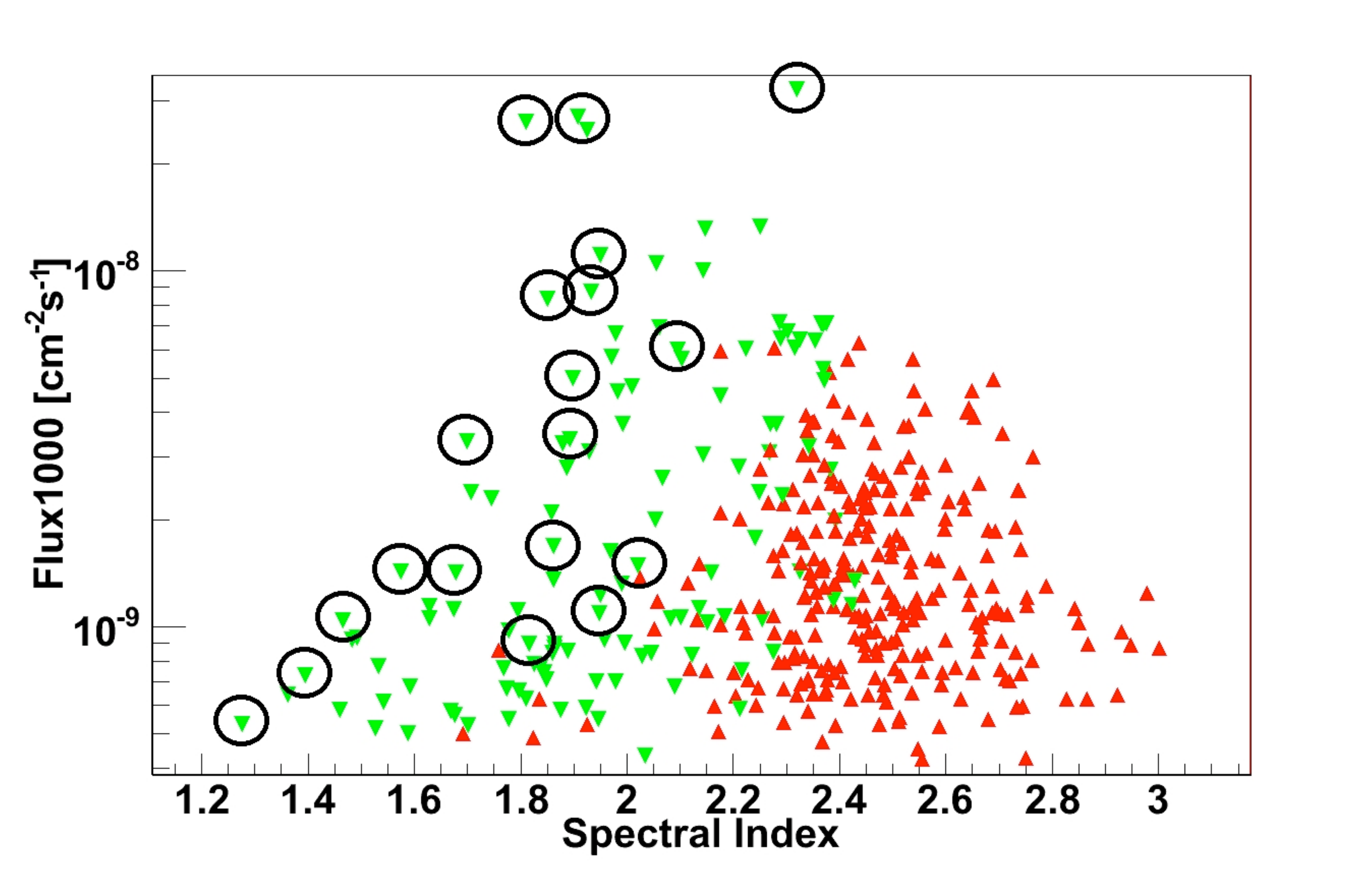}
  \caption{
{\it Fermi} Spectral index and {\it Fermi} flux F1000 for AGN that would
exceed the 5$\sigma$ level in less than 50 hours with CTA. The
points correspond to detections
(green filled inverted triangles) and non-detections (red filled
triangles) respectively. Large black empty circles denote
AGN detected by current ground-based Cherenkov
observatories. CTA candidate array E was used to produce this
plot.}
  \label{sources}
\end{figure*}

\section{Results}

Using the outlined recipe, we analyzed the complete 400 sources in 
the AGN sample extracted from the {\it Fermi} 1FGL catalog.   
We obtain $> 120$ AGN detections, or approximately three times 
the current sample of VHE AGN at the time of this writing. 
In order to visualize our results in context of 
current detections, Figure \ref{sources} shows 
the {\it Fermi} flux (F1000) and spectral index for 
expected $> 5\sigma$ detections with CTA.
Also shown are VHE AGN detected with the current generation
of instruments. 
It is obvious that CTA will be most efficient for 
hard sources ($\Gamma<2$) and will reveal the AGN population
beyond the tip of the iceberg of current detections. 
For softer sources, CTA will allow us to detect 
quiescent sources that are
currently only accessible during prominent flares.

It 
must be noted that the artificial break introduced at 100 GeV for 
hard sources might restrict the number of detections. Furthermore, 
we have left out nearly 200 BL Lacs with unknown redshift listed in
the 1FGL. These are typically $\Gamma<2$ sources where CTA is most efficient.
As a result, 120 AGN must be considered a conservative 
lower limit
for  the AGN detection rate with CTA. Therefore, 
these forecasts are quite encouraging overall.

\subsection{Redshift limits}

Apart from increasing the actual number of AGN detected, CTA should 
increase dramatically the
number of AGN visible at high redshifts. 
The most distant quiescent AGN predicted with our code
is at $ z = 1.8$. However,
certain conditions  could push that limit to even higher redshift. 
In particular, our estimations indicate that, certain flaring 
FSRQs with a 
gamma-ray flux increase of a factor of 10 and moderate 
spectral hardening 
$\Delta \Gamma = 0.3$ could potentially produce a detection out to $z = 2.9$.
Alternatively, a fraction of well-studied {\it Fermi} BL Lacs could be 
detected out to $z = 1.2$. The main difficulty in 
elucidating the BL Lac population will be
actually obtaining direct redshift measurements from 
featureless optical/UV 
spectra. A possible redshift workaround might come from a direct
measure of the EBL shape that could allow to set an upper limit for
the redshift \cite{raue}.

\section{CTA Survey Capabilities}

Even though CTA will not be able to match the
{\it Fermi}-LAT in cadence, 
its wide field of view (FoV) capabilities 
will allow easy access to wide portions of the sky. Analyzing
our results,  
we find that a dedicated pointed survey with the
CTA Observatory  should detect an excess of 
120 sources in less than a year for sources observed for 
a maximum of 50 hours. Over a year and assuming a 5 degree 
effective FoV, the large number of pointings should 
produce an initial sky survey covering  5\% to 7\%
of the sky by default. Although these pointings  will not reach 
equivalent flux levels, there is a relative high 
probability of finding interesting sources 
from serendipitous detections (see for example \cite{MAGIC_IC310}).

An alternative survey approach could  
select a continuous region of interest and image it deeply 
(at least  5 hours  per pointing). 
Depending on the specific details such a wide field
 survey could cover 400 to 4000 square
degree stripes per site per year. 
This deeper survey could aim for well-mapped areas in other wavelengths 
to allow for multifrequency analyses, 
and be oriented to both: a) probe the faint end of AGN population and 
b) guide the design of subsequent observations.

\section{Conclusions}

The work presented here illustrates the tremendous CTA capabilities
compared to current ground-based Cherenkov instruments. An
excess of 120 AGN is expected as a first order approximation. 
This number could swell to close to 300 AGN considering that 
at least 200 {\it Fermi} 1FGL
BL Lacs without redshift have not been included 
in this analysis. Thus, our results
must be considered as a conservative lower limit. Further 
increases are expected from
the discovery of AGN that eluded {\it Fermi} detection with gamma-ray 
emission peaking in the CTA energy range. 

The highest AGN redshift detected should be pushed from current $z = 0.5$ to 
approximately $z \approx 2$. 
It could  even reach higher values if one takes into account 
flaring FSRQs or if ongoing spectroscopic surveys manage 
to constrain the redshifts of more distant {\it Fermi} BL Lacs. 
All the results presented 
are preliminary as further refinements 
to CTA array configurations and improvements to analysis tools
are introduced. 
It is important to bear in mind that there are important
caveats in our calculations: uncertainties in the spectral parameters,
limitations in the CTA effective area 
calculations, and the EBL model. 
But regardless of the actual outcome, it is clear that 
we look forward to a 
 densely populated sky in the TeV range with CTA.
This will open new opportunities to AGN studies
including dedicated multiwavelength campaigns and
studies of flares short time-scales. 
The next logical step for this work will come soon with the release 
of the Second {\it Fermi} LAT Catalog (2FGL). With improved spectral 
fitting and a larger number of sources, the 2FGL should allow us 
to better understand the CTA capabilities. Additional 
redshift constraints of BL Lac objects will help us to better account for
the full AGN population. Finally, improvements on 
EBL absorption models should produce more 
detailed results.

\section{Acknowledgments}
We thank Catherine Boisson, Helene Sol and Andreas Zech for organizing
a very interesting workshop. 
We are indebted to the CTA Monte Carlo Work
Package for their exceptional work during the design study.  
We gratefully acknowledge support from the agencies and
organizations listed in this page: 
\href{http://www.cta-observatory.org/?q=node/22}
{http://www.cta-observatory.org/?q=node/22}.
The authors 
acknowledge the support of the Spanish Government under project code
FPA2010-22056-C06-06.  N.M. gratefully acknowledges support from
the Ram\'on y Cajal Fellowship program.


\begin{thebibliography}{99999}
\bibitem{design} The CTA Consortium 2011, Exp. Astron., 32, 193
\bibitem{fermi} Abdo, A. A. et al., 2010a, \emph{ApJS}, 188, 405
\bibitem{Konrad} Bernl\"{o}hr, K., 2008, \emph{American Institute of Physics Conference 
Series}, 1085, 874
\bibitem{Fermi_AGN_cat} Abdo, A. A. et al., 2010b, 
\emph{ApJ}, 715, 429
\bibitem{CGRaBS} Healey, S. E. et al., 2008, 
\emph{ApJ}, 175, 97
\bibitem{Roma-BZCAT} Massaro, E. et al., 2009,
\emph{ApJ}, 495, 691
\bibitem{zhang} Zhang, J. et al., 2011, 
 \emph{ApJ}, submitted 
[arXiv:1108.0607v1]
\bibitem{Gould} Gould, R. J. \& Schreder, G.,  1996, 
\emph{Physical Review Letters}, 16, 252
\bibitem{EBL_Franceschini} Franceschini, A.  et al., 2008,
\emph{A\&A}, 487, 837
\bibitem{Li&Ma} Li, T. \& Ma, Y., 1983, 
\emph{ApJ}, 272, 317
\bibitem{raue} Raue, M. \& Mazin, D., 2011, \emph{APh}, 34, 245
\bibitem{MAGIC_IC310} Aleksi{\'c}, J. et al., 2010, 
\emph{ApJ}, 723, L207
 
	
\end{thebibliography}
\end{document}